\documentclass[11pt,aps,prd,showpacs,byrevtex,nofootinbib]{revtex4}
\usepackage{bm,amsmath,amssymb,amsfonts,latexsym,psfrag}
\usepackage{graphics}
\usepackage{graphicx}

\begin{document}

\title{\Large Casimir-Polder intermolecular  forces
in minimal length theories}
\author{O. Panella}
\affiliation{INFN Sezione di Perugia, Via Pascoli, Perugia Italy}

\date{August 8, 2007}

\begin{abstract}
\textit{Generalized uncertainty relations are known to provide a
minimal length $\hbar\sqrt{\beta}$.   The effect of such minimal
length in the Casimir-Polder interactions between neutral atoms
(molecules) is studied. The first order correction term  in the
minimal uncertainty parameter is derived and found to describe an
attractive potential scaling as $r^{-9}$ as opposed to the well
known $r^{-7}$ long range retarded potential.}
\end{abstract}
\maketitle
\def\constant{\kappa}
\def\Si{\textrm{Si}}
\section{Introduction}
The complete understanding of the properties of the quantum vacuum
is a point of central importance to both fundamental and applied
physics~\cite{wilczek}. A celebrated  mechanical manifestation of
the fluctuations of the electromagnetic field around its zero point
(vacuum) configuration is the well known Casimir
effect~\cite{casimir} whereby an attractive force arises between two
metallic neutral plates separated by a distance $d$~\footnote{See
however ref.~\protect\cite{jaffe} for a different viewpoint which
claims that there is no need to invoke vacuum fluctuations to
explain the Casimir effect.}. The Casimir effect has recently
received a lot of attention mainly because of the following reasons:
$i$) advances in the experimental techniques leading to precision
measurements~\cite{lamoreaux,bressi1} and to the possibility of
measuring the dynamic Casimir effect~\cite{bressi2}; $ii)$ increased
importance of the Casimir forces in the field of
micro-electro-mechanical systems (MEMS)~\cite{roukes} and
nano-devices~\cite{capasso}; $iii)$ relevance to physics beyond the
standard model of particle physics, such as, for example,
hypothetical extra-dimensional models~\cite{long}. The present work
is a contribution to the last of the above points and tries to make
a connection between the Casimir effect  and models which are
generally refereed to as minimal length theories.

The concept of a minimal length arises naturally in quantum gravity.
Indeed, when trying to resolve small distances, higher energies are
needed which eventually  will affect the structure of space-time via
their gravitational effects. Clearly this type of effect should
occur at an energy scale of the order of the Planck mass. While it
is generally stated that the incorporation of gravity within quantum
field theory spoils the renormalizability of the latter, the fact
that quantum gravity could play an important role in the suppression
of infinities in quantum field theories has been pointed out long
ago~\cite{salam}.

Other models have been discussed in the literature which are also
related to a minimal length such as for example extra-dimensional
theories, and/or non-commutative quantum field
theories~\cite{arkani,refsonXD,nouicer2}. It is hoped that the
energy (or equivalently length) scale associated to these models
might turn out to be accessible experimentally in the next
generation of high-energy accelerator such as the Cern LHC or the
International Linear Collider.

 Although not trivial to prove it is natural to expect that a minimal distance
should correspond at the quantum level to a minimal uncertainty. On
the other hand string theories have been shown to predict the
existence of a \emph{minimal length} in the form of an effective
minimal uncertainty $(\Delta x)_{min}$~\cite{stringminimal}. This
has triggered the study of quantum mechanical models based on
generalized commutation relations such as:
\begin{equation}
\label{gcr}
[\hat{x},\hat{p}] = i\hbar
\left(1 +\beta \hat{p}^2 
\right)\, ,
\end{equation}
which lead to a generalized uncertainty principle (GUP) which in
turn provides a minimal uncertainty. In particular in
Ref.~\cite{kempf1} an Hilbert space representation of the
generalized uncertainty relations is constructed explicitly in a one
dimensional model, as well as in higher dimensions, identifying the
states which realize the maximal localization (i.e. the minimal
uncertainty) in position space. An important aspect of these
theories is that the eigen-states of the position operator \emph{are
no longer physical states} and so one is forced to introduce the
so-called quasi-position representation, which consists in
projecting the states onto the set of maximally localized states,
instead of using the standard position representation obtained
projecting the state vectors on the eigen-states of the position
operator. Thus momentum states which are normally represented by
plane waves $(2\pi)^{-3/2}\exp[{({i}/{\hbar})\bm{p}\cdot\bm{x}}]=
\langle\bm{x}|\bm{p}\rangle $ are now replaced by the set of
functions $ \psi^{ML}_{\bm{p}}(\bm{x})  = \langle
\psi^{ML}_{\bm{x}}|\bm{p}\rangle $, where $ \psi^{ML}_{\bm{x}}$ are
the maximally localized states (minimal uncertainty) around an
average position $\bm{x}$. The particular form of these states
depends on the number of dimensions and on the specific model
considered. These type of models are also connected to
non-commutative quantum field theories.

In Ref.~\cite{nouicer} the author studies the standard
Casimir effect of two perfectly conducting plates at a distance $d$
computing the correction term arising
within a  model based on generalized uncertainty relations:
\begin{equation}
\label{result_nouicer}
{\cal U}_{Casimir} =  - \frac{\hbar\, c\, \pi^2}{720\, d^3}
\left[ 1  -\pi^2\, \frac{ 99504 }{56855}\,
\left(\frac{\hbar\sqrt{\beta}}{d}\right)^2 \right] .
\end{equation}
It is worth to be noted that the correction term due to the
minimal length $(\Delta x)_{min} \approx \hbar\sqrt\beta$ scales with
a different power law ($d^{-6}$) and it is repulsive.
This result is used
to obtain an upper bound on the minimal length which turns out  to be
$(\Delta x)_{min} \le 150$ nm.

In this work we take up the above scenario of generalized uncertainty
relations and compute the correction term to the
intermolecular Casimir-Polder interactions induced by a minimal length,
expecting
to find results similar to those of Ref.~\cite{nouicer}
 (c.f. Eq.~\ref{result_nouicer}). In particular,
by performing a quantization procedure of the electro-magnetic field
as in Ref.~\cite{nouicer}, it will
be shown that if a minimal length exists in nature the retarded Casimir-Polder
interactions between neutral atoms and molecules will acquire,
in addition to the standard $r^{-7}$ interaction,
a corrective term which scales as $r^{-9}$.
However, as opposed to the plate-plate case, c.f. Eq.~(\ref{result_nouicer}),
the new term has the same sign of the standard QED result,
i.e. it describes an  attractive interaction.

The reminder of the paper is organized as follows. In Section
\ref{gupsection} we discuss the generalized uncertainty relations
and define a set of maximally localized states; in section \ref{QED}
we discuss the quantization of the electromagnetic field in the
presence of a minimal length and,  in section
\ref{CasimirPolder}, we derive the corrections to the Casimir-Polder
intermolecular interactions due to a minimal length. Finally in
section \ref{conclusions} we present the conclusions.

\section{Generalized Uncertainty Principle}
\label{gupsection}
Let us consider the generalized commutation relations of Eq.~\ref{gcr},
for simplicity in one dimension.
From these modified commutation relations one derives the
generalized uncertainty principle (GUP)~\cite{kempf1}
\begin{equation}\label{GUP}
\triangle x \triangle p \geq \frac{\hbar}{2} [1+ \beta  (\triangle
p)^2 + \langle \hat{p} \rangle^2 ] \qquad \beta > 0 \, ,
\end{equation}
which is found to be related to a minimal
length~\cite{nouicer,kempf1}. Indeed with a simple minimization procedure
it is easily verified~\cite{kempf1} that
Eq.~(\ref{GUP}) implies an absolute minimal uncertainty $(\triangle x)_{min}
=\hbar\sqrt{\beta}$, which is obtained for those states such that
$\langle \hat{p}\rangle =0$ and
$ (\Delta p)^2 = \langle \hat{p}^2 \rangle =1/{\beta}$.
An Heisenberg algebra that satisfies the generalized commutation
relations of Eq.~(\ref{gcr}) is represented on momentum space wave functions
by:
\begin{eqnarray}
\label{p_rep}
\hat{p}\, \psi(p) &=&  p\, \psi(p)\, ,\\
\hat{x}\, \psi(p) &=& i \hbar\,(1+\beta p^2)\, \partial_p\, \psi(p)\, .
\end{eqnarray}
In Ref.~\cite{kempf1} the eigenstates of the position operator have
been shown to be non physical because their uncertainty in the
position (which vanishes) is smaller than the absolute minimal
uncertainty (or minimal length). One then can look for a  set of
maximally localized states for which the uncertainty in the position
is minimal (i.e. equal to the minimal length)~\cite{kempf2}. The
procedure proposed by Kempf, Mangano and Mann (KMM)~\cite{kempf1} to
find the maximally localized states consists in minimizing the value
$(\Delta x)^2_\psi$ between those states which realize the
generalized uncertainty principle in Eq.~(\ref{GUP}) with the equality
sign (\emph{i.e. a squeezed state}):
\begin{equation}
\label{squeezed}
(\Delta {x})_\psi (\Delta {p})_\psi = \frac{1}{2}\,
\left|\,\langle [\hat{x},\hat{p}] \rangle_\psi \right|\, .
\end{equation}
In the adopted momentum space representation Eq.~(\ref{squeezed}) takes the form of a
differential equation:
\begin{equation}
\label{diffeq}
\left[i\hbar (1+\beta p^2)\, \partial_p -\langle \hat{x}\rangle +i\hbar
\frac{1+\beta (\Delta p)^2 +\beta \langle p \rangle}{2(\Delta p)^2}
\, (\hat{p}
-\langle\hat{p}\rangle) \right] \, \psi(p) = 0\, ,
\end{equation}
which admits the normalized solution
$\left(\langle \hat{x} \rangle = \xi,\, \langle \hat{p} \rangle = 0,\,
\Delta p = 1/\sqrt{\beta}\right)$:
\begin{equation}
\label{MLstatesonedim} \psi^{ML}_\xi (p) =\sqrt{\frac{2\sqrt{\beta}
}{\pi}}\, (1+\beta p^2)^{-\frac{1}{2}} \, \exp\left[-i \frac{\xi
\arctan (\sqrt{\beta}p)}{\hbar \sqrt{\beta}}  \right]\, .
\end{equation}
As opposed  to ordinary quantum mechanics it turns out that such states,
in addition to being normalizable, are  of finite energy and
no longer ``orthogonal''  i.e. their closure relation involves
a finite function instead of a
Dirac-$\delta$ distribution ($\delta(\xi -\xi')$):
\begin{equation}
\langle \psi^{ML}_{\xi'} | \psi^{ML}_\xi \rangle = \frac{1}{\pi}
\left[\frac{\xi-\xi'}{2\hbar \sqrt{\beta}} -
\left( \frac{\xi-\xi'}{2\hbar \sqrt{\beta}}\right)^3    \right]^{-1}
\, \sin\left(\pi \frac{\xi-\xi'}{2\hbar \sqrt{\beta}} \right)\, .
\end{equation}
 \\

\noindent{\sf{\textbf{Extension to higher dimensions.}}} In
Ref.\cite{kempf2} the above ideas have been extended to a number $n$
of arbitrary (spatial) dimensions. An important fact  to recall is
that there is no unique extension of Eq.~\ref{gcr} in more than one
dimension. Indeed in order to preserve translational and rotational
invariance generalized commutation relations must take the
form~\cite{kempf1}:
\begin{equation}
[\hat{x}_i,\hat{p}_j] = i\hbar \left[ f(\hat{p}^2) \delta_{ij}
+g(\hat{p}^2)\hat{p}_i \hat{p}_j \right]\, , \qquad i,j =1, \dots ,n\, .
\end{equation}
The functions $f(\hat{p}^2)$ and $g(\hat{p}^2)$ are not completely
arbitrary. Relations between them can be found by imposing
translational and rotational invariance.  In the following we take
up the standard choice of Ref.~\cite{kempf2} which has become rather
popular and has been the object of many phenomenological studies:
\begin{equation}
\label{model} f(\hat{p}^2)=
\frac{\beta\hat{p}^2}{\sqrt{1+2\beta\hat{p}^2}-1}\, ,\qquad \qquad
g(\hat{p}^2)=\beta\, .
\end{equation}
A spectral representation can then be found such that:
\begin{equation}
[\hat{x}_i, \hat{z}_j]= i \hbar \delta_{ij} \, , \qquad     \hat{x}_i =
{i}{\hbar}\partial_{z_i}\, .
\end{equation}

The KMM procedure consists in minimizing the position uncertainty
within the set of squeezed states, see Ref.~\cite{kempf1,kempf2} for
details, and yields the following \emph{maximally localized states}
around a mean position $\bm{\xi}$ (in the spectral representation):
\begin{equation}
\label{MLstates} \psi^{ML}_{\bm{\xi}}( \bm{z}) = N \left(1-\beta
\frac{z^2}{2}\right)^{\alpha/2} \exp\left[-\frac{i}{\hbar}
\bm{z}\cdot\bm{\xi}\right]\, ,
\end{equation}
where  $\alpha=1+\sqrt{1+n/2}\approx 2.58$ (in three spatial
dimensions) is a numerical constant that characterizes the maximally
localized states in the KMM approach. On the other hand when the
number of dimensions $ n \ge 1$ the generalized uncertainty
relations are not unique and different models (actually an infinite
number of them) may be implemented~\cite{kempf2} by choosing
different functions $f(p^2)$ and/or $g(p^2)$ (c.f. Eq.~\ref{model})
which will yield in general \emph{different} maximally localized
states which will not contain at all the numerical constant
$\alpha$. The \emph{minimal} position uncertainty for the KMM
maximally localized states is~\cite{detournay}:
\begin{equation}
(\Delta x)_{min} = \hbar\sqrt{\beta}\, \left[ \frac{n}{8}\,
\left(1+\sqrt{1+\frac{n}{2}}\right) \left(\frac{n+2}{\sqrt{1+n/2}}+2
\right)\right]^{1/2}\,= \hbar\sqrt{\beta}\,\sqrt{\frac{n}{4}}\,
\alpha\, ,
\end{equation}
which in the case of three spatial dimensions gives $(\Delta
x)_{min} =  2.23533\, \hbar\sqrt{\beta}\,$.

\begin{figure}
  \scalebox{0.85}{\includegraphics{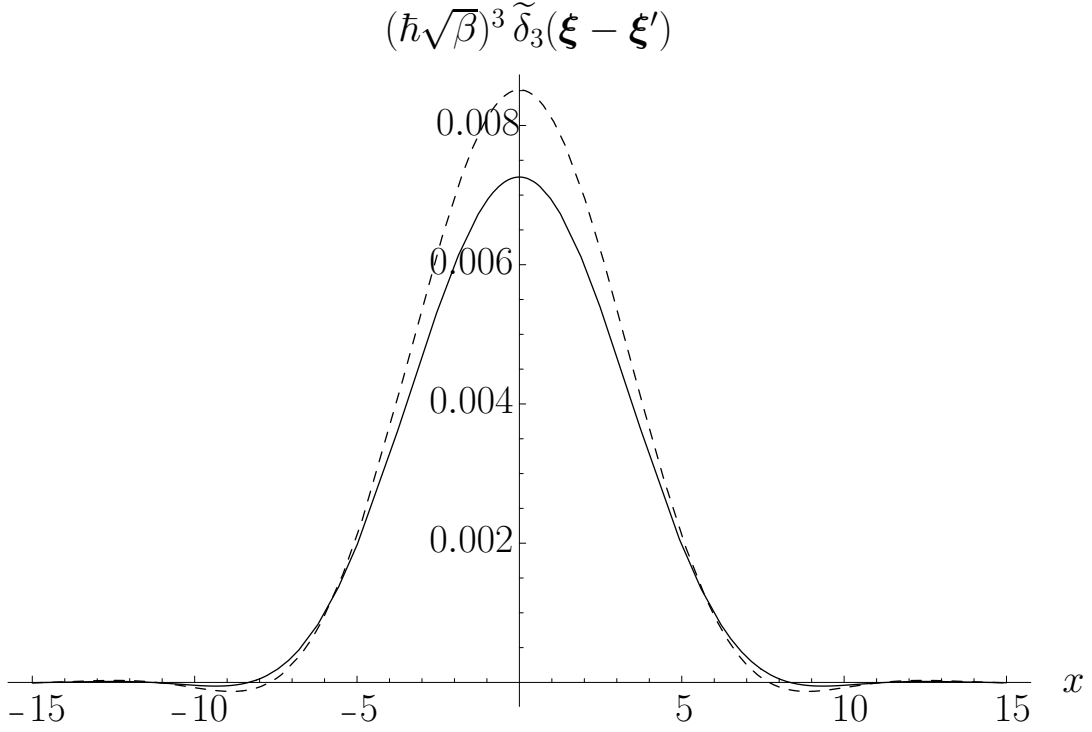}}
  \caption{Plot of the \emph{regular} closure function
  $(\hbar\sqrt{\beta})^3\, \widetilde{\delta}_3(\bm{\xi} -\bm{\xi}')$
  along  an arbitrary direction of the vector $\bm{\xi} -\bm{\xi}' $
  (both positive and negative),
  versus the dimension-less
  quantity $x= \sqrt{2} |\bm{\xi} -\bm{\xi}'| /(\hbar\sqrt{\beta})$;
  (a) the dashed line is the result obtained (\protect{Eq.~\ref{closurekmm}})
  within the KMM
  procedure of Ref.~\cite{kempf2}, with $\alpha=1+\sqrt{1+3/2}$
  corresponding to three spatial dimensions ($n=3$); (b) the solid line
  is the result [Eq.~\ref{closuredgs} of this work] of the calculation performed with
  the (DGS) approach of Ref.~\protect\cite{detournay}.
  This plot should not mislead the reader. Both functions satisfy the integral
  relation $ \int \, d^3\bm{\xi}'\, \widetilde{\delta}_3(\bm{\xi} -\bm{\xi}') =1 $.
  The importance of numerical difference between the two functions, apparently
   inconsistent with this constraint, is reduced,
  when doing the integral, by the $x^2$ factor from the differential
  $d^3(\bm{\xi-\xi'})$, and indeed both closure functions do integrate to 1,
  as has been checked both analytically and numerically. }
  \label{closurefig}
\end{figure}

 The states given in
Eq.~(\ref{MLstates}), are normalizable states whose closure relation
reads:
\begin{equation}
\label{closure} \langle \psi^{ML}_{(\bm{\xi})} |
\psi^{ML}_{(\bm{\xi}^\prime)} \rangle = \widetilde{\delta}_n
(\bm{\xi} -\bm{\xi}') = N^2 \int_{|\bm{z}|\leq
\sqrt{\frac{2}{\beta}}} d^n\bm{z}
\left(1-\beta\frac{z^2}{2}\right)^{\alpha}\,
\exp\left[-\frac{i}{\hbar} \bm{z}\cdot(\bm{\xi} -
\bm{\xi}^\prime)\right]\, ,
\end{equation}
which turns out be a finite function for  finite $\beta$. The
normalization constant $N$ can be chosen so that
$\widetilde{\delta}_n(\bm{\xi} -\bm{\xi}') $  reduces to the
Dirac-$\delta$ distribution $\delta^n (\bm{\xi} -\bm{\xi}')$ in the
limit of zero minimal length ($\beta \to 0$) i.e. $N=
(2\pi\hbar)^{-n/2}$. Explicitly one has (for three spatial
dimensions, $n=3$):
\begin{equation}
\label{closurekmm} \widetilde{\delta}_3 (\bm{\xi} -\bm{\xi}^\prime)
= \Gamma (1+\alpha)\, \frac{2^\alpha}{\pi^{3/2}
(\hbar\sqrt{\beta})^3}\, \left(\frac{\sqrt{2}|\bm{\xi}
-\bm{\xi}^\prime|}{\hbar\sqrt{\beta}} \right)^{-(3/2+\alpha)} \,
J_{3/2+\alpha} \left(\frac{\sqrt{2}|\bm{\xi}
-\bm{\xi}^\prime|}{\hbar\sqrt{\beta}} \right)\, .
\end{equation}
\noindent In Fig.~\ref{closurefig} the closure function
$\widetilde{\delta}_3$ is plotted against its dimensionless
argument. It can be easily verified that it satisfies the relation
$\int d^3\bm{\xi} \,\widetilde{\delta}_3 (\bm{\xi} -\bm{\xi}^\prime)
= 1 $.

\noindent For the purpose of deriving the first order correction in
the minimal length of the Casimir-Polder forces it will prove useful
to expand the closure function, c.f. Eq.~(\ref{closure}), in powers of
$\beta$ and in terms of the Dirac-$\delta$ distribution. In order to
properly do this one must go to the momentum representation (see
appendix A for details) and one finds for the present model and
within the KMM approach:
\begin{equation}
\widetilde{\delta}_n(\bm{r} -\bm{r}')
   \approx  \left[ 1
 + (\hbar\sqrt{\beta})^2 \, \frac{\constant}{2}\, \nabla^2_{\bm{r}}+
 \ldots \right] \delta^n (\bm{r} -\bm{r}^{\prime}) \, .
 \label{MLclosure}
\end{equation}
where $\constant = 2+\alpha$. We take the previous equation as the
definition of the numerical constant $\constant$ whose value will in
general depend on the model considered~\footnote{We emphasize that
$\constant$ is a \emph{numerical constant} and not a free parameter.
Within the KMM procedure for obtaining the maximally localized
states its numerical value (within the model considered here) is
fixed by the number of spatial dimensions only.}.\\ It should be
noted that the KMM procedure to construct maximal localization
states works properly only for the generalized commutation relations
of Eq.~(\ref{gcr}). Detournay, Gabriel and Spindel
(DGS)~\cite{detournay} have proposed a better definition of
maximally localized states, which is based on a minimization
procedure on the subset of all physical states, and not just on the
subset of squeezed states, and turns out to be suitable for more
general commutation relations than those considered in
Eq.~(\ref{gcr}). It should be realized that adopting this method
(Ref.~\cite{detournay}) would only slightly affect our conclusions,
namely Eq.~(\ref{MLclosure}). The maximally localized states are in
this approach given by Bessel functions. In the end result, c.f.
Eq.~(\ref{MLclosure}),  we can anticipate that expanding the closure
function in powers of the minimal length $(\hbar\sqrt\beta)$ it would
merely amount to a change in the value of the numerical constant
$\constant$ which can however be expected to be always of order
unity.

Let us discuss in further detail this point. In particular we would
like to show that adopting the DGS procedure, see
Ref.~\cite{detournay}, the resulting regular closure function
$\widetilde{\delta}_3(\bm{r} -\bm{r}')$ although represented by a
different analytic expression is numerically very close to that
computed in the KMM procedure. Furthermore its expansion in powers
of $\beta$ is as given in Eq.~(\ref{MLclosure}) but with a slightly
lower value of the numerical constant $\constant$. In
Ref.~\cite{detournay} the authors show that the maximally localized
states corresponding to the particular model being discussed here
(c.f. Eq.~(\ref{model})) are given (in the spectral representation) in
terms of Bessel functions:
\begin{equation}
\psi_{\bm{\xi}}^{ML}(\bm{z})= C_n \, \frac{J_\nu(\mu z)}{z^\nu}\,
\exp\left[-\frac{i}{\hbar}
\bm{\xi}\cdot\bm{z}\right]\, ,
\end{equation}
where  $z= |\bm{z}|$; $\nu=\frac{n}{2}-1$,
$\mu=\pi\sqrt{\beta}/\sqrt{2}$ and the normalization constant $C_n$
depends again on the number of dimensions that we are considering.
These states are characterized by a \emph{minimal} uncertainty which
is given by:
\begin{equation}
(\Delta x)_{min} = \hbar \mu = \frac{j_{\nu,1}}{\sqrt{2}}\,
\hbar\sqrt{\beta}\, ,
\end{equation}
$j_{\nu,1}$ being the first zero of the Bessel function $J_\nu(x)$.
For $n=3$ we have  $\nu= 1/2$ and $j_{\nu,1}=\pi$ so that the
minimal length derived in the DGS procedure is $ (\Delta x)_{min} =
2.22144\, \hbar \sqrt\beta $ which is indeed smaller than the
minimal length derived in the KMM approach\footnote{Recall that the
minimization in the DGS procedure is carried through within the set
of all physical states while in the KMM approach  it is restricted
to the squeezed states.}. With these states the closure function
becomes:
\begin{equation}
\label{closure2} \langle \psi^{ML}_{(\bm{\xi})} |
\psi^{ML}_{(\bm{\xi}^\prime)} \rangle = \widetilde{\delta}_n
(\bm{\xi} -\bm{\xi}') = C_n^2 \int_{|\bm{z}|\leq
\sqrt{\frac{2}{\beta}}} d^n\bm{z} \, \frac{\left[J_\nu(\mu
z)\right]^2}{z^{2\nu}}\, \exp\left[-\frac{i}{\hbar}
\bm{z}\cdot(\bm{\xi} - \bm{\xi}^\prime)\right]\, .
\end{equation}
Using the well known expansion for the Bessel functions:
\begin{equation}
\label{BesselJ} J_\nu(t) =
\frac{1}{\Gamma(\nu+1)}\,\left(\frac{t}{2}\right)^\nu\, \left[
1-\frac{1}{\nu+1} \, \left(\frac{t}{2}\right)^2 + ....\right]\, ,
\end{equation}
the constant $C_n$ can be chosen so that  in the limit of a
vanishing minimal length $\beta \rightarrow 0$ (or $\mu \rightarrow
0$) the regular closure function $\widetilde{\delta}_n (\bm{\xi}
-\bm{\xi}') $ reduces to the Dirac-$\delta$ distribution ${\delta}^n
(\bm{\xi} -\bm{\xi}') $:
\begin{equation}C_n=
\frac{\Gamma(n/2)}{(2\pi\hbar)^{n/2}}
\left(\frac{2}{\mu}\right)^{n/2-1}\, .
\end{equation}
In addition it is possible to derive an expansion of the closure function of the type in Eq.~\ref{MLclosure}  also within the DGS
procedure. One derives the following
expansion in powers of $\beta$ (see Appendix A for details):
\begin{equation}\langle \psi^{ML}_{(\bm{\xi})} |
\psi^{ML}_{(\bm{\xi}^\prime)} \rangle = \widetilde{\delta}_n
(\bm{\xi} -\bm{\xi}') \approx  \left[ 1
 + \frac{2+(j_{\nu,1})^2/n}{2}\,{(\hbar\sqrt\beta)^2} \, \nabla^2_{\bm{\xi}}+
 \ldots \right] \delta^3 (\bm{\xi} -\bm{\xi}^{\prime}) \, ,
 \label{MLclosure2}
\end{equation}
thereby defining, also in the DGS procedure, a numerical constant
$\constant$, namely: $\displaystyle\constant =2+(j_{\nu,1})^2/n $.

On the other hand by analytically carrying through the integration
in Eq.~(\ref{closure2}) one finds the closed expression (when $n=3$):
\begin{equation}
\label{closuredgs} \widetilde{\delta}_3 (\bm{\xi} -\bm{\xi}')
=\frac{\sqrt{2}}{4\pi^4}\, \frac{1}{(\hbar\sqrt{\beta})^3}\,
\frac{2\Si(x)-\Si(x-2\pi)-\Si(x+2\pi)}{x}\, , \end{equation} where again
$x= {\sqrt{2}|\bm{\xi}-\bm{\xi}'|}/(\hbar\sqrt{\beta})$ and
$\Si(x)=\int_0^x\, dt\, \sin(t)/t$ is the sine-integral function.
One again can verify that $\widetilde{\delta}_3 (\bm{\xi}
-\bm{\xi}')$ satisfies the relation $\int d^3\bm{\xi}
\,\widetilde{\delta}_3 (\bm{\xi} -\bm{\xi}^\prime) = 1 $.
Figure~\ref{closurefig} shows a comparison of the closure functions
derived in the KMM and DGS procedure. Table~I summarizes some results within the two approaches for the model of the present work (model I).
\begin{table}\label{table}\caption{Within the rotationally invariant
model (Model I) of Eq.~(\ref{model}) (see
Ref.\protect\cite{kempf1,kempf2}) we compare the numerical constant
$\constant$ and the minimal uncertainty as discussed in the text for
both the KMM and DGS procedure of obtaining the maximally localized
states. Recall that the constant $\alpha = 1 +\sqrt{1+n/2}$ appears
in the maximally localized states of the KMM procedure. We also give
the same details for the direct product model (Model II) used in
Ref.~\cite{nouicer}, which is reviewed for the reader's benefit in
Appendix A.}
\begin{ruledtabular}
\begin{tabular}{lcccc}
& $\constant$& $\constant\, (n=3)$ &$\displaystyle\frac{(\Delta
x)_{min}}{\hbar\sqrt\beta} \,(n) $ & $\displaystyle\frac{(\Delta
x)_{min}}{\hbar\sqrt\beta}\, (n=3)$ \vspace{0.1cm}\\\hline
\vspace{0.1cm} Model I (KMM)&$\displaystyle{2+\alpha}$&$
2+1+\sqrt{5/2}\approx 4.58114
$&$\displaystyle{\sqrt{\frac{n}{4}}\,\alpha}$&
$\displaystyle{\sqrt{\frac{3}{4}}\,\alpha \approx 2.2353}$\\
 Model I (DGS)&$\displaystyle{2+\frac{
(j_{n/2-1,1})^2}{n}}$&$ \displaystyle{2+\frac{\pi^2}{3}\approx
5.28987}$& $\displaystyle\frac{j_{n/2-1,1}}{\sqrt{2}}$&
$\displaystyle{\frac{\pi}{\sqrt{2}} \approx 2.22144}$\vspace{0.2cm}\\
 Model II &10/3&10/3&$\sqrt{n}$&$\sqrt{3}$
\end{tabular}
\end{ruledtabular}
\end{table}

The above discussion shows in particular that \emph{even within the
same model} adopting the KMM procedure or the more appropriate DGS
method (where the minimization of the uncertainty is done over all
physical states) will produce \emph{slightly} different values of
the numerical constant $\constant$, thereby justifying the use of
the widespread  KMM approach at least within this particular model.

\section{QED in the temporal gauge}
\label{QED} In order to approach the quantization procedure  in
presence of a minimal length we review briefly the derivation of the
temporal gauge $\left[ A^0(\bm{r},t)=c\varphi(\bm{r},t)=0 \right]$
propagator in \emph{standard} QED. The defining equation for the
photon propagator in QED is:
\begin{equation}\label{green}
D^{ij}(\bm{r},t;\bm{r}',t') = \frac{i}{\hbar }\,\langle 0| T \left[
A^i(\bm{r},t)\, A^j(\bm{r}',t') \right] |0\rangle ,
\end{equation}
from which, applying through the classical equation of motion for the
fields $A_i(\bm{r},t) $ (Maxwell equations) in the temporal gauge, it
is easily verified that:
\begin{equation} \left[
\left(\frac{1}{c^2}\frac{\partial^2}{\partial t^2} -\nabla^2\right)
\delta^\ell_{\phantom{l}m} -\nabla^\ell\nabla_m \right] D^{mk}
(\bm{r},t;{\bm{r}}',t') = - \frac{i}{\hbar c} \delta(t-t') \langle
0| \left[ E^\ell(\bm{r},t), A^k(\bm{r'},t')\right]|0\rangle\, ,
\label{mixedGreen}
\end{equation}
where the electric field operator $\bm{E}(\bm{r},t) = -(1/c)\partial
\bm{A}(\bm{r},t) /\partial t $ has been introduced. Canonical
quantization prescribes then the equal-time commutation
relations~\cite{zuber}:
\begin{equation}
 \left[E^\ell(\bm{r},t) , A^k(\bm{r}',t) \right] = 4\pi\, i \hbar c\,
 \delta^{\ell k}\,
 {\delta}^3 (\bm{r} -\bm{r}') ,
 \label{eqtcr}
\end{equation}
and the differential equation for the QED photon propagator becomes:
\begin{equation} \left[
\left(\frac{1}{c^2}\frac{\partial^2}{\partial t^2} -\nabla^2\right)
\delta^\ell_{\phantom{l}m} -\nabla^\ell\nabla_m \right] D^{mk}
(\bm{r}-{\bm{r}}';t-t') = 4 \pi \delta^{\ell k}\delta(t-t')
{\delta}^3 (\bm{r} -\bm{r}')\, . \label{fullpropQED}
\end{equation}
The solution is found by going into a Fourier representation
\begin{equation}
D^{rj} (\bm{k};\omega) = \int dt\, d^3\bm{r} \, e^{i\omega t}\,
e^{-i\bm{k}\cdot \bm{r}}D^{rj}(\bm{r},t)\, ,\end{equation} and is given
by the well known result:
\begin{equation}
\label{DK} D^{rj}(\bm{k},\omega) = -\frac{4\pi}{(\omega/c)^2 -k^2
+i0} \, \left[ \delta^{rj} -\frac{k^rk^j}{(\omega/c)^2}\right]\, .
\end{equation}
As regards the calculation of the effect of a minimal length in the
Casimir-Polder  intermolecular interactions,
following~\cite{landau}, it will be convenient to use the propagator
in the mixed representation:
\begin{equation}
D^{ij}(\omega,\bm{r}) = \int_{-\infty}^{+\infty}\, dt\, e^{i\omega
t}\, D^{ij}(t,\bm{r})\, ,
\end{equation}
which is  found upon integrating in $\bm{k}$-space Eq.~(\ref{DK}):
\begin{equation}
D^{ij}(\omega,\bm{r}) = \int \, \frac{d^3 \bm{k}}{(2\pi)^3}\,
e^{i\bm{k}\cdot \bm{r}} \, D^{ij} (\bm{k},\omega) \, .
\end{equation}
It is easily verified that $ D^{ij}(\omega,\bm{r}) $ satisties the
following differential equation:
\begin{equation}
\label{GREEN_QED} \left[ -\left(\frac{\omega^2}{c^2} +
\nabla^2\right) \delta^\ell_{\phantom{\ell}m} -\nabla^\ell\nabla_m
\right] D^{mk} (\omega,\bm{r}) = 4\pi\, \delta^{\ell k}\,
\delta^3(\bm{r})\, ,
\end{equation}
which is solved by~\cite{landau}:
\begin{equation}
\label{explicit_propagator} D^{ij}(\omega,\bm{r}) = \left[
\delta^{ij} \left(
1+\frac{ic}{|\omega|r}-\frac{c^2}{\omega^2r^2}\right) +
\frac{x^ix^k}{r^2} \left(
\frac{3c^2}{\omega^2r^2}-\frac{3ic}{|\omega|r} -1 \right) \right]
\frac{e^{i|\omega|r/c}}{r}\,\,  .
\end{equation}
\\

\noindent {\textsf{\textbf{Quantization in minimal length
theories}}}\\
Let us now discuss the quantization procedure of the
electro-magnetic field in a quantum world with a minimal length. We
shall proceed following the scheme adopted in~\cite{nouicer}. In
this case the procedure of canonical quantization gets modified
because it turns out that the equal-time commutation relations of
the fields, see Eq.~(\ref{eqtcr}), are different. Indeed now
instead of expanding the field operators over plane waves
(\emph{position representation wave functions of momentum states})
one is forced to expand the fields over a set of maximally localized
states $|\psi^{ML}_{\bm{r}}\rangle$ of average position $\bm{r}$
(\emph{quasi-position representation of momentum states}):
\begin{eqnarray}
A^i(\bm{r},t) &=& \sum_\lambda \int
\frac{d^3\bm{p}}{(2\pi\hbar)^3}\left(\frac{2\pi\hbar
c^2}{\omega_{\bm{p}}}\right)^{1/2} \, \left[
a^{\phantom{\dagger}}\!({\bm{p},\lambda})\,\varepsilon^i (\bm{p},\lambda) \, \langle
\psi^{ML}_{\bm{r}}|\bm{p}\rangle\, e^{-i\omega_{\bm{p}}t}\right.\cr
& &\phantom{xxxxxxxxxxxxxxxxxxxxxxxxxxxx}\left.
+a^\dagger({\bm{p},\lambda})\,(\varepsilon^i (\bm{p},\lambda))^*\,
\langle \bm{p}\,|\psi^{ML}_{\bm{r}}\,\rangle
\,e^{+i\omega_{\bm{p}}t} \right]\, ,
\end{eqnarray}
introducing creation and annihilation operators which do satisfy the
usual commutation relations,
\begin{equation}
[a({\bm{p},\lambda}),a^\dagger({\bm{p}',\lambda'})] = (2\pi)^3\,
\delta^{\lambda,\lambda'}\,\delta^3(\bm{p}-\bm{p}')\, ,\end{equation}
with all other commutators vanishing (\emph{recall that momentum
coordinates are commuting in the model which is being discussed
here}). The equal time commutation relations for the fields, c.f.
Eq.~(\ref{eqtcr}), are then easily found to be modified to:
\begin{eqnarray}
\label{modified_eqcr}
 \left[E^\ell(\bm{r},t) , A^k(\bm{r}',t) \right] &=&
 4\pi\, i\hbar c\,
 \delta^{\ell k}\, \langle \psi^{ML}_{\bm{r}}|\psi^{ML}_{\bm{r}'}\rangle
 \nonumber \\
 &=& 4\pi\, i\hbar c\,
 \delta^{\ell k}\,
 \widetilde{\delta}_3 (\bm{r} -\bm{r}')\, ,
\end{eqnarray}
so that, comparing with Eq.~(\ref{GREEN_QED}), the photon propagator in
the presence of a minimal length,
$\widetilde{D}^{ij}(\omega,\bm{r})$, is found from:
\begin{equation}
\left[ -\left(\frac{\omega^2}{c^2} + \nabla^2\right) \delta^{\ell}_
{\phantom{\ell}m} -\nabla^\ell\nabla_m \right] \widetilde{D}^{mk}
(\omega,\bm{r}) = 4\pi\,  {\delta}^{\ell k}\,
\widetilde{\delta}_3(\bm{r}) \label{GREEN_GUP}\, .
\end{equation}
This approach to field quantization bears some similarity to that
discussed in Ref.~\cite{spallucci} where within a model of
non-commutative space-time the real scalar field is defined as a
mean value over coherent states. The non-commutativity is reflected
by a modification of the Fourier transform replacing ordinary plane
waves by gaussian wave-packets. We remark that in
Ref.\cite{spallucci} the authors derive for the Green function a
differential equation which is similar to our Eq.~(\ref{GREEN_GUP})
where the three-dimensional Dirac $\delta$-function is replaced by a
gaussian dumped by the non-commutative parameter which plays the
role of the parameter $\beta$ of this work.\\   Being $
\widetilde{\delta}_3(\bm{r})$  a regular function, it turns out that
the standard QED photon propagator ${D}_{rj} (\omega,\bm{r})$ (from
Eq.~(\ref{GREEN_QED})) is the  Green function of Eq.~(\ref{GREEN_GUP}),
the differential equation describing the photon propagator in
\emph{minimal length} QED, $\widetilde{D}^{ij} (\omega,\bm{r})$.
Therefore it follows the central result of this work. \\\emph{In
minimal length QED the photon propagator in the mixed representation
is given by a convolution of the regular closure function
$\widetilde{\delta}_3(\bm{r})$ with the standard QED photon
propagator}:
\begin{equation} \label{convolution}
\widetilde{D}^{ik} (\omega,\bm{r}) = \int d^3\bm{r}'
D^{ik}(\omega,\bm{r}-\bm{r}') \, \widetilde{\delta}_3 (\bm{r}' )\, .
\end{equation}
From Eq.~\ref{MLclosure} one can obtain, by means of partial
integration, an expansion of the photon propagator in terms of the
minimal length $\hbar \sqrt{\beta}$:
\begin{equation}
\widetilde{D}^{ik} (\omega,\bm{r})\approx D^{ik}(\omega,\bm{r})
+(\hbar\sqrt{\beta})^2\,\frac{\constant}{2}\, \nabla^2
D^{ik}(\omega,\bm{r}) \, .
\end{equation}
Using the explicit form of the QED temporal gauge propagator given
in Eq.~(\ref{explicit_propagator}) it is easily verified that:
\begin{equation}
\nabla^2 {D}^{ik} (\omega,\bm{r}) = -\frac{\omega^2}{c^2} {D}^{ik}
(\omega,\bm{r}) \, ,
\end{equation}
so that we finally get:
\begin{equation}\label{approxD}
\widetilde{{D}}^{ik} (\omega,\bm{r}) \approx
\left[1-\frac{\constant}{2}\,(\hbar\sqrt{\beta})^2\,
\frac{\omega^2}{c^2} \right] {{D}}^{ik} (\omega,\bm{r})\, .
\end{equation}
A remark is in order at this point. It should be clear to the reader
that the above approach to field quantization within a minimal
length model breaks Lorentz invariance. 
This fact appears as well in other extensions of
the standard model (SM) such as non-commutative quantum field
theory, being always one of the major sources of debates between
different authors. It should be noted that in Ref.~\cite{spallucci}
the question of Lorentz covariance is discussed and, within their
non-commutative 2D model, successfully addressed. While completing
this study the author became aware of a recent work~\cite{quesne}
where a relativistic generalization of the Kempf algebra is
proposed. Probably, based on this new deformed algebra, it will be
possible to define a Lorentz covariant field quantization which
would account for a \emph{fundamental minimal length}. Certainly
this point deserves further study, but goes beyond the scope of the
present work.

\section{Casimir-Polder intermolecular interactions}
\label{CasimirPolder} The interaction of two neutral atoms
(molecules) at rest in $\bm{r}_1$ and $\bm{r}_2$ and with electric
dipole moments $\bm{d}_1$ and $\bm{d}_2$ is described by the
operator:
\begin{equation}
V = -\bm{E}(\bm{r}_1,t)\cdot \bm{d}_1(t) -\bm{E}(\bm{r}_2,t)\cdot
\bm{d}_2(t)
\end{equation}
where $\bm{E}(\bm{r},t)$ is the operator describing the electric
field. It is well known that, employing standard perturbation
methods of quantum field theory, the interaction of the neutral
atoms (or molecules) at distances $r=|\bm{r}_1-\bm{r}_2|$, which are
large compared to the atomic and/or  molecular dimensions $ a$: $r
\gg a$, \footnote{In this field theoretic approach the distance $r$
can otherwise be either much smaller (\emph{short distances}) than
the characteristic wavelength $\lambda_0$, of the spectra of the
interacting atoms (or molecules), comparable or much larger than
$\lambda_0$, (\emph{long distances}). } is described in terms of
$(i)$ the dynamic polarizability tensor $\alpha^{ik}(\omega)$ of the
atoms:
\begin{equation}
 \alpha^{ik}
(\omega)=\frac{i}{\hbar}\,\int_{-\infty}^{+\infty}\, {d\tau}\,
\langle 0 |T\left[\,d^i(\tau)\, , \, d^k(0)\, \right]|0\rangle\,
e^{+i\omega \tau} \, ,
\end{equation}
 and $(ii)$ the photon propagator in
the mixed representation $D^{ik}(\omega,\bm{r})$~\cite{landau}.
For the photon propagator we will use in the following the dyadic (tensor)
notation ${\sf{D}}(\omega, \bm{r})$.
Assuming that the polarizability tensors of the two atoms are
isotropic $ \alpha^{ik}_{(1,2)}(\omega)  = \delta^{ik}\,
\alpha_{(1,2)}(\omega)$, one obtains:
\begin{equation}
\alpha(\omega) = \frac{1}{3}\, \sum_n |d_{0n}|^2 \, \left (
\frac{1}{\omega_{n0}-\omega -i0^+}+\frac{1}{\omega_{n0}+\omega
-i0^+}\right)\, ,
\end{equation}
and the final expression for the interaction potential can be cast
as:
\begin{equation}
\label{potential} U(r) =
\frac{i\hbar}{2c^4}\int_{-\infty}^{+\infty}\,
\frac{d\omega}{2\pi}\,\omega^4\, \alpha_1(\omega)\,
\alpha_2(\omega)\, {\tt{Tr}} \Big[{\sf{D}}(\omega,\bm{r})\cdot
{\sf{D}}(\omega,\bm{r}) \Big]\, .
\end{equation}
 When a minimal length is present in the above
expression one should replace the photon propagator $ {\sf{D}}
(\omega, \bm{r})$ with the modified propagator $
\widetilde{{\sf{D}}} (\omega, \bm{r})$. Using the approximate
expression in Eq.~(\ref{approxD}) one finds:
\begin{equation}
{\tt{Tr}} \left[\widetilde{\sf{D}}(\omega,\bm{r})\cdot
\widetilde{\sf{D}}(\omega,\bm{r}) \right] \approx
\left[1-{\constant}\,(\hbar\sqrt{\beta})^2\, \frac{\omega^2}{c^2}
\right]\, {\tt{Tr}} \Big[{\sf{D}}(\omega,\bm{r})\cdot
{\sf{D}}(\omega,\bm{r}) \Big]\, ,
\end{equation}
so that the interaction potential is given by the usual standard
model result with an additional term describing the effects of the
minimal length:
\begin{equation}
\label{correction} \delta U(r) = -\frac{i}{2}\,\frac{\hbar}{c^6}
(\hbar\sqrt{\beta})^2\, \constant\,\int_{-\infty}^{+\infty}\,
\frac{d\omega}{2\pi}\,\omega^6\, \alpha_1(\omega)\,
\alpha_2(\omega)\, {\tt{Tr}} \Big[{\sf{D}}(\omega,\bm{r})\cdot
{\sf{D}}(\omega,\bm{r}) \Big]\, .
\end{equation}
By using the explicit expression in Eq.(\ref{explicit_propagator}) it
is easily found :
\begin{equation}
{\tt{Tr}} \Big[{\sf{D}}(\omega,\bm{r})\cdot {\sf{D}}(\omega,\bm{r})
\Big] = 2 \left[1+ \frac{2ic}{|\omega| r}-\frac{5c^2}{\omega^2 r^2}
-\frac{6ic^3}{|\omega|^3r^3}+\frac{3c^4}{\omega^4 r^4}\right]
\frac{e^{2i|\omega| r/c}}{r^2} \label{trDD}\, .
\end{equation}
\\
The general expressions in Eq.~(\ref{potential}) and
Eq.~(\ref{correction}) can be simplified in the limiting cases of
short distances ($a  \ll r \ll \lambda_0$) and long distances ($ r
\gg \lambda_0 $).\\
\textbf{\textsf{Short distances}}\\
When $ r \ll \lambda_0 $, in the integral of Eq.~(\ref{correction}) are important the values
of $\omega \sim \omega_0 \sim c/\lambda_0$ so that $\omega r/c \ll
1$ and in Eq.~(\ref{trDD}) one is allowed to keep only the last term
and to approximate the exponential with 1. Performing the
straightforward computations one then finds an interaction:
\begin{equation}
\label{correctionCP} U(r) = - \frac{2}{3r^6} \sum_{n,n'}
\frac{|d_{n0}|^2\, |d_{n'0}|^2}{\hbar(\omega_{n0}+\omega_{n'0})}
\left[1- \constant\, (\hbar\sqrt{\beta})^2
\,\left(\frac{\omega_{n0}^2 +\omega_{n'0}^2}{c^2}\right) \right]\, ,
\end{equation}
which scales as $r^{-6}$ and is composed of two terms: (i) the well
known London type potential~\cite{landau} and the correction term
which is due to the minimal length. In this regime of short
distances the minimal length correction affects only the strength of
the interaction and does not change its power law.\\
 \textbf{\textsf{Long distances}}\\ \noindent In the limit of
\emph{long distances} $r \gg \lambda_0$ in the integral of Eq.~(\ref{correction}) are
important only the values of $ \omega \lesssim c/r \ll \omega_0$.
When $\omega \gtrsim  \omega_0$, the strongly oscillating complex
exponential will suppress the integral. It is then possible to
substitute the dynamic polarizabilities $\alpha_{1,2}(\omega)$  with
their static values $\alpha_1(0)$ and $\alpha_2(0)$. We then find a
correction of order $(\hbar\sqrt{\beta}/r)^2$ to the well known
$r^{-7}$ Casimir-Polder interaction:
\begin{equation}
\delta U(r) = - \frac{129}{\phantom{|}8\pi}\,\constant\,
\left(\frac{\hbar\sqrt{\beta}}{r}\right)^2\, \frac{\hbar\, c\,
\alpha_1(0)\alpha_2(0)}{r^7}\, .
\end{equation}
Thus, in theories with a minimal length,  a correction term arises
in the Casimir-Polder atomic and molecular interactions at large
distances. The correction has the same sign (\emph{attractive
potential}) but scales with a different power law ($r^{-9}$)
relative to the QED result:
\begin{equation}
{\cal U}_{Casimir-Polder}(r) \approx - \frac{23}{4\pi} \,
\frac{\hbar\, c\, \alpha_1(0)\alpha_2(0)}{r^7}
\left[1+\frac{129}{56}\, \constant \,
\left(\frac{\hbar\sqrt{\beta}}{r}\right)^2\right] \, .
\label{retardedinteraction}
\end{equation}
This might be compared with the result of Ref.~\cite{nouicer} c.f.
Eq.~(\ref{result_nouicer}) which describes the Casimir energy of the
plate-plate system. One can notice that while in the plate-plate
case the minimal length correction derived in Ref.~\cite{nouicer} is
\emph{repulsive} and opposite to the standard result, in the case
treated here of the interaction of two neutral atoms or molecules
the minimal length correction is \emph{attractive} i.e. of the same
sign of the standard Casimir-Polder result. We should however be
very careful in comparing the calculation of the Casimir effect of
Ref.~\cite{nouicer} with the present one. Indeed in
Ref.~\cite{nouicer} a different  model of generalized  uncertainty
relations is explicitly taken up.  As opposed to our Eq.~(\ref{model})
the model proposed in Ref.~\cite{nouicer} given by $f(p^2) = 1+\beta
p^2$ and $ g(p^2)=0 $ is the naive generalization of the
one-dimensional model. Such model is known~\cite{matsuo} to be
inconsistent with the KMM construction of maximally localized
states. Due to the resulting non-commutativity of the coordinates
the resulting $n$ differential equations (corresponding to Eq.~(\ref{diffeq})) of
the squeezed states cannot be solved simultaneously~\cite{matsuo}. A
different way of extending the one-dimensional GUP of Eq.~(\ref{gcr})
is to take a direct product of it. This so-called direct product
model (model II) clearly breaks rotational invariance as opposed to
the model of this paper (model I). The maximally localized states
are obtained by taking the product of the one-dimensional states,
see Eq.~(\ref{MLstatesonedim}), along the different dimensions. This
model is therefore characterized by different maximally localized
states, which will in turn produce different closure functions
which, \emph{when expanded in powers of $\beta $}, will provide
different values for the numerical constant $\constant$ as defined
by Eq.~(\ref{MLclosure}) (though we expect it to be always of order
unity). In the appendix we provide details of the closure function
within this model and deduce the corresponding value of $\constant$
which turns out to be $\constant = 10/3$. We may therefore conclude
that the Casimir-Polder minimal length correction within the direct
product model is also attractive. That the Casimir effect for
parallel plates within the direct product model (model II) as
calculated in Ref.\cite{nouicer} turns out to be repulsive may be
due to the fact that the closure function there appears to have been
computed not with the maximally localized states, but with the
formal position eigenstates which are not physical states. Setting
this point definitively calls for a detailed analysis of the Casimir
effect using the \emph{proper} maximally localized states within
either model I or II  and
 goes beyond the scope of the present work.

The so-called Casimir-Polder force, for the atom-atom configuration,
was first derived in 1948~\cite{casimirpolder}, but it was not
measured definitively until 1993~\cite{sukenik}, by looking at the
deflection of an atomic beam passing through two parallel plates.
For earlier measurements related to the the atom-plane
configuration, see~\cite{sparnay}.  There has been then a consistent
renewal of interest with an increase of the measurement's precision
as in Ref.~\cite{harber} where this tiny interaction between a
neutral system and a surface is reported. The authors employed a
cloud of ultra-cold atoms in a Bose-Einstein Condensate (BEC) state.
The range of distances explored was between 6 and 10 microns. See
the recent review~\cite{onofrio} for more details about the first
generation of experiments, both in the Casimir effect and the related Casimir-Polder interactions, and for later developments and future directions.

It seems that  a better direction to explore the Casimir-Polder
interactions experimentally is to consider the plate-sphere
(plate-atom) configuration. All the experimental advances  have been
achieved for this case~\cite{onofrio}. This suggests the direction
where the present theoretical work could develop: the calculation of
the Casimir-Polder (plate-atom) interaction corrective term within a
minimal length theory.

\section{Discussion and Conclusions}
\label{conclusions} We have studied the implications of models based
on generalized commutation relations \emph{i.e.  with a minimal
length} in the Casimir-Polder intermolecular interactions. The
calculation is done following standard perturbation theoretical
methods of quantum field theory and in particular the approach of I.
E. Dzjalo\u{s}inskij as illustrated in quantum field theory
textbooks~\cite{landau} is used. Here the interaction energy of two
neutral atoms and/or molecules is related to the dynamic
polarizability tensors of the two neutral bodies and to the photon
Green function (propagator). The computation of the correction term
due to the minimal length is thus carried out by discussing the QED
photon propagator in presence of a minimal length. Quantization of
the electro-magnetic field in the temporal gauge is performed in
analogy to the canonical quantization procedure, the essential point being that the field operators $A^i(\bm{r},t)$ instead of being
decomposed over a complete set of plane waves (momentum
eigenfunctions) are now decomposed over a complete set of maximally
localized states. This approach has also been followed by the author of Ref.~\cite{nouicer} in deriving the Casimir potential energy of
the plate-plate system in the presence of a minimal length c.f.
Eq.~(\ref{result_nouicer}). The main point is that the equal-time
commutation relations of the field operators instead of being given
in terms of a Dirac $\delta$-function, c.f. Eq.~(\ref{eqtcr}),  are
now expressed by a finite regular function c.f.
Eq.~(\ref{modified_eqcr}), and the photon propagator in minimal
length QED is given by a convolution  of the standard QED propagator
with the regular function $\widetilde{\delta}_3(\bm{r}-\bm{r}')$. In
order to compute the lowest order correction term 
the minimal length QED propagator is related to that of standard QED
by performing an expansion in the minimal length parameter $ \hbar
\sqrt{\beta}$.

We have derived a corrective term to the (long distance) retarded
Casimir-Polder interaction of two neutral atoms separated by a
distance $r \gg \lambda_0$, finding, c.f.
Eq.~(\ref{retardedinteraction}), a new interaction term whose
potential scales like $r^{-9}$ as opposed to the standard $r^{-7}$
result.

Clearly should the minimal length $(\Delta x)_{min} =\hbar
\sqrt{\beta}$ be of the order of the Planck length $L_P$ the
observability of this effect would be out of question. In
Ref.~\cite{brau} the study of the harmonic oscillator and the
Hydrogen atom allowed to derive an upper bound on the minimal length
by comparing with precision measurements on hydrogenic atoms and for
electrons trapped in strong magnetic fields. The upper bound
obtained is:
\begin{equation}
\label{upperbound} (\Delta x)_{min} =\hbar\sqrt{\beta}< 10^{-1}
\text{fm}\, .
\end{equation}
Assuming that the constant $\beta$ appearing in the deformed
Heisenberg algebra is a universal constant, this upper bound would
presumably preclude a possible observation with Casimir-Polder
interaction measurements with accessible distances $r$ which are
typically in the range $80$ nm  $\leq r \leq$ $10$
$\mu$m ~\cite{onofrio}. Thus we conclude that taking into account the upper bound
of Ref.~\cite{brau}, the term derived in this work, c.f.
Eq.~(\ref{retardedinteraction}), would provide a correction whose
relative strength we expect to be in the range:
\begin{equation}
10^{-22}   \lesssim \frac{\delta U}{U}
=O\left[\left(\frac{\hbar\sqrt{\beta}}{r}\right)^2\right] \lesssim
10^{-18}\, ,
\end{equation}
which is clearly beyond any foreseeable  improvement in the
precision of measurements of Casimir-Polder interactions.  As
discussed in Ref.~\cite{brau2} the rather strong upper bound in
Eq.~(\ref{upperbound}) could  be avoided by assuming that the
parameter $\beta$ is not a universal constant and could vary from a
system to another depending for example on the energy content of the
system (the mass of the particle for instance) or the strength of
some interaction. Indeed in Ref.~\cite{brau2} by making this
hypothesis the authors, through a comparison with the experimental
results for ultra-cold neutron energy levels in a gravitational
quantum well (GRANIT experiment)~\cite{granit1,granit2,granit3}
derive a relaxed upper bound to the minimal length which turns out
to be of the order of a few nano-meters [$(\Delta x)_{min} < 2.41 $
nm]. Even if this situation should apply, the relative strength of
the corrective term will  fall in the range:
\begin{equation}
10^{-8}   \lesssim \frac{\delta U}{U}
=O\left[\left(\frac{\hbar\sqrt{\beta}}{r}\right)^2\right] \lesssim
10^{-4}\, ,
\end{equation}
which still is out of the reach of current experiments in the
Casimir-Polder interactions, whose precision is typically of
$10^{-2}\,  \text{or}\,  1\%$~\cite{onofrio}. With an increased
experimental sensitivity to smaller length scales, \emph{of the
order of a few nano-meters}, and better experimental accuracy
possible perhaps in the near future, the calculations described in
the present work could find interesting applications and would
provide a valuable tool to study the intimate structure of
space-time.

\begin{acknowledgments}

The author wishes to acknowledge partial support from the
\textsc{Fondazione Cassa di Risparmio di Spoleto (Italy)}.
\end{acknowledgments}

\appendix

\begin{section}{Details of the Closure functions}
In this appendix we give details of the series expansion in powers
of $\beta$ of the closure function both in the model taken up in
this work and in the one considered in Ref.~\cite{nouicer}.
In particular we show that in all considered cases the following
relation  holds with different values of $\constant$:
\begin{equation}
\widetilde{\delta}_n(\bm{\xi}-\bm{\xi}') = \left[ 1+
\frac{\constant}{2}\, (\hbar\sqrt\beta)^2\, \nabla^2_{\bm{\xi}}
+\text{O}(\beta^2) \right]\, \delta^n(\bm{\xi}-\bm{\xi}')
\end{equation}
It turns out that in order to get the correct series expansion in
powers of $\beta$ of the closure function one must resort to the
unbounded momentum representation, while both Eq.~\ref{closure} and
Eq.~\ref{closure2}, based on the spectral representation (compact
space), can only reproduce correctly the first term of the expansion
corresponding to $\beta=0$.  Let us consider this in detail within: $(i)$ the rotational invariant model object of this
work, (model I) detailed both within the KMM and DGS procedure  and $(ii)$ the direct product  model of Ref.~\cite{nouicer}, (model II).\vspace{0.2cm} \\
\textsf{Rotational Invariant Model (Model I)}\\ In
Eq.(\ref{closure}) we first perform a change of variables from the
spectral representation to the momentum representation:
\begin{equation}
z_i= p_i \frac{\sqrt{1+2\beta p^2} -1 }{\beta
p^2}\, , \qquad\text{with}\qquad {{p}}^2={\bm{p}}\cdot{\bm{p}} = \sum_i^n (p_i)^2\, .
\end{equation}The Jacobian matrix is:
\[
\frac{\partial z_i}{\partial p_j}= \delta_{ij}a +2p_ip_j b\, ,
\]
where:
\[
a=\frac{\sqrt{1+2\beta p^2} -1 }{\beta p^2} , \qquad b=
\frac{-1-\beta p^2 +\sqrt{1+2\beta p^2}}{\beta p^4 \sqrt{1+2\beta
p^2}}\, .
\]
The Jacobian of the transformation $J$ turns out to be
given by:
\[
J= det \left| \frac{\partial z_i}{\partial p_j}\right| = a^{n-1}
(a+2bp^2) = \left(\frac{\sqrt{1+2\beta p^2} -1 }{\beta p^2}\right)^n
\frac{1}{\sqrt{1+2\beta p^2}}\,.
\]
\vspace{0.0cm}\\\noindent
\underline{a) KMM procedure for the maximally localized states}\\
Eq.(\ref{closure}) becomes:
\begin{equation}
\widetilde{\delta}_n(\bm{\xi}-\bm{\xi}') = \int \, \frac{d^n
\bm{p}}{(2\pi \hbar)^n}  \left( \frac{\sqrt{1+2\beta p^2} -1 }{\beta
p^2} \right)^{n+\alpha}\, \frac{1}{\sqrt{1+2\beta p^2}} \,
\exp\left[-\frac{i}{\hbar} (\bm{\xi}-\bm{\xi}')\cdot \bm{p}
\frac{\sqrt{1+2\beta p^2} -1 }{\beta p^2}\right]\, .
\end{equation}
We remark that the integration is unbounded. The integrand can be
expanded in a power series around $\beta =0$:
\begin{equation}
\label{closurexpansion} \widetilde{\delta}_n(\bm{\xi}-\bm{\xi}') =
\int \, \frac{d^n \bm{p}}{(2\pi \hbar)^n} \, \left[ 1 -\beta
\left(\frac{n+2+\alpha}{2}p^2
-\frac{i}{2\hbar}(\bm{\xi}-\bm{\xi}')\cdot \bm{p}\, p^2 \right)
+\text{O}(\beta^2) \right]\,\exp\left[-\frac{i}{\hbar}
(\bm{\xi}-\bm{\xi}')\cdot \bm{p}\right]\, .
\end{equation}
In the above expression one can make the substitution $\bm{p} \to
i\hbar \nabla_{\bm{\xi}}$ and thereby perform the momentum
integration obtaining:
\begin{equation}
\widetilde{\delta}_n(\bm{\xi}-\bm{\xi}') = \left[ 1+
\frac{2+\alpha}{2}\, (\hbar\sqrt\beta)^2\, \nabla^2_{\bm{\xi}}
+\text{O}(\beta^2) \right]\, \delta^n(\bm{\xi}-\bm{\xi}')\, ,
\end{equation}
thus defining for this case the value of the numerical constant
$\constant$ to be given by $\constant = 2+\alpha$. We note that the
last expression has been obtained after using identities (due to
partial integration) valid for distributions such as:
\begin{equation}
\left[(\bm{\xi}-\bm{\xi}')\cdot \nabla_{\bm{\xi}}\right] \,
\nabla^2_{\bm{\xi}} \delta^n (\bm{\xi}-\bm{\xi}') = -n\,
\nabla^2_{\bm{\xi}} \delta^n (\bm{\xi}-\bm{\xi}')\, .
\end{equation}\\
\noindent \underline{{b) DGS procedure for the maximally localized states}}\\
Using the maximally localized states of the DGS procedure one finds
in the momentum representation:
\begin{eqnarray}
\widetilde{\delta}_n(\bm{\xi}-\bm{\xi}') &=& C_n^2 \int  d^n \bm{p}
\left( \frac{\sqrt{1+2\beta p^2} -1 }{\beta p^2} \right)^{n}
\frac{1}{\sqrt{1+2\beta p^2}} \,\left.\frac{\left[J_\nu(\mu
z)\right]^2}{z^{2\nu}}\right|_{z=\sqrt{\bm{z}\cdot\bm{z}}=
\frac{\sqrt{1+2\beta p^2}-1}{\beta\sqrt{p^2}}}\times \nonumber\\
&&\phantom{xxxxxxx}\,\exp\left[-\frac{i}{\hbar}
(\bm{\xi}-\bm{\xi}')\cdot \bm{p} \frac{\sqrt{1+2\beta p^2} -1
}{\beta p^2}\right]\, ,
\end{eqnarray}
where again the momentum integration is unbounded and the integrand
admits a well defined series expansion in powers of $\beta$. Using
Eq.\ref{BesselJ} one finds similarly to Eq.~\ref{closurexpansion}:
\begin{eqnarray}
\label{closurexpansionb} \widetilde{\delta}_n(\bm{\xi}-\bm{\xi}')
&=& \int \, \frac{d^n \bm{p}}{(2\pi \hbar)^n} \, \left[ 1 -\beta
\left(\frac{n+2+(j_{\nu,1})^2/{n}}{2}p^2
-\frac{i}{2\hbar}(\bm{\xi}-\bm{\xi}')\cdot \bm{p}\, p^2 \right)
+\text{O}(\beta^2) \right]\,\times\nonumber
\\ &&\phantom{xxxxxxxxxxxxxxxxxxxxxxxxiixxxxxxxxxxxx}
\exp\left[-\frac{i}{\hbar} (\bm{\xi}-\bm{\xi}')\cdot
\bm{p}\right]\, ,
\end{eqnarray}
and performing the momentum integration as before we end up with:
\begin{equation}
\widetilde{\delta}_n(\bm{\xi}-\bm{\xi}') = \left[ 1+
\frac{2+(j_{\nu,1})^2/{n}}{2}\, (\hbar\sqrt\beta)^2\,
\nabla^2_{\bm{\xi}} +\text{O}(\beta^2) \right]\,
\delta^n(\bm{\xi}-\bm{\xi}')\, ,
\end{equation}
therefore defining for the DGS procedure a value of the numerical
constant $\constant$  given by $\constant = 2 +(j_{\nu,1})^2/n$.
\\
\noindent {\textsf{Direct Product Model (Model II)}}\\
We note that while in Ref.~\cite{nouicer} the author announces   to
study the model:
\begin{equation}
[\hat{x_i},\hat{p_j}] = i \hbar (1+\beta \hat{\bm{p}}^2 )\delta_{ij}
\, , \qquad i=1,\dots ,n \, , \qquad \hat{\bm{p}}^2=\hat{\bm{p}}\cdot\hat{\bm{p}} = \sum_i^n (\hat{p}_i)^2\, ,
\end{equation}
then he uses, as maximally localized states,
the direct product of the one-dimensional states.

More precisely the proper definition of the direct product  model
(model II) is~\cite{matsuo}:
\begin{equation}
[\hat{x}_i,\hat{p}_j] = i \hbar (1+\beta \hat{p}_i^2 )\delta_{ij}\, ,
\qquad i=1,\dots ,n \, .
\end{equation}
In this model the closure function is given by:
\begin{equation}
\label{ADP} \widetilde{\delta}_n(\bm{\xi}-\bm{\xi}') = \int \,
\frac{d^n \bm{p}}{(2\pi \hbar)^n}
\prod_{j=1}^n\,\frac{1}{\left(1+\beta p_j^2\right)^2}\,
{\displaystyle\exp\left[+\frac{i}{\hbar}(\xi_j-\xi_j')
\frac{\text{Arctan}(\sqrt{\beta}p_j)}{\displaystyle\sqrt{\beta}}\right]}
\, ,\end{equation}
which factorizes in the product of $n$ identical integrals:
\begin{eqnarray}
\widetilde{\delta}_n(\bm{\xi}-\bm{\xi}') &=& \prod_{j=1}^n\,
\int_{-\infty}^{+\infty}\, \frac{dp}{2\pi\hbar}
\,\frac{1}{\left(1+\beta
p^2\right)^2}\,\exp\left[+\frac{i}{\hbar}(\xi_j-\xi_j')
\,\frac{\text{Arctan}(\sqrt{\beta}p)}
{\sqrt{\beta}}\right]\\
&=&\prod_{j=1}^n\, \frac{1}{\hbar\sqrt{\beta}}\,
\frac{1}{4}\frac{\sin{[x_j]}}{x_j-x_j^3/\pi^2}\, ,
\end{eqnarray}
with  $ x_j={\pi(\xi_j-\xi_j')}/(2\hbar\sqrt{\beta})$, while in
Ref.~\cite{nouicer} it is found:
\[\widetilde{\delta}_n(\bm{\xi}-\bm{\xi}') =
\prod_{i=1}^n\,\frac{\sin\left(\frac{\pi(\xi_j
-\xi_j')}{2\hbar\sqrt{\beta}}\right)}{\pi(\xi_j-\xi_j')}=
\prod_{i=1}^n\, \frac{1}{\hbar\sqrt\beta}\,\frac{1}{2}\,
\frac{\sin[x_j]}{x_j}\, .
\] We remark that in Ref.~\cite{nouicer}, in addition to some
notational inconsistencies, the closure function appears to have
been calculated improperly, as the $(1+\beta p^2)$ factor appears in the denominator only with one power, while it should be squared, as
one such factor is due to the change in the momentum measure and the other is from the square of the wave functions of the maximally
localized states, see Ref.~\cite{kempf1}. By performing a series
expansion in powers of $\beta$ of the integrand in Eq.~(\ref{ADP}) we end up with the following:
\begin{equation}
\widetilde{\delta}_n(\bm{\xi}-\bm{\xi}') = \left[1 +
\frac{5}{3}(\hbar \sqrt{\beta})^2 \nabla^2_{\bm{\xi}} +
\text{O}(\beta^2)\right]\, \delta_n(\bm{\xi}-\bm{\xi}')\, ,
\end{equation}
which defines for this model $\constant=10/3$, as reported in Table
I.

Let us remark that within this model, given by the direct product of
the one-dimensional algebra discussed in section II, the KMM
procedure gives the exact maximally localized states and so  there
is no need to distinguish between the KMM and DGS
procedure~\cite{detournay}.

It was however important to check that the sign of the constant
$\constant$ is the same as in model I. Thus even within the direct
product model the Casimir-Polder correction due to the minimal
length turns out to be attractive.

\end{section}
 \printendnotes

%
\end{document}